\documentclass[traditabstract]{aa} 
\usepackage{graphicx}
\usepackage{natbib}
\bibpunct{(}{)}{;}{a}{}{,}
\usepackage{txfonts}
\begin{document}
   \title{{\it Herschel}/PACS far-infrared photometry of two $z>4$
     quasars\thanks{{\it Herschel} is an ESA space observatory with science instruments 
       provided by European-led Principal Investigator consortia and with important 
       participation from NASA}}

   \author{C. Leipski\inst{\ref{inst1}}
     \and
     K. Meisenheimer\inst{\ref{inst1}} 
     \and
     U. Klaas\inst{\ref{inst1}}
     \and
     F. Walter\inst{\ref{inst1}}     
     \and
     M. Nielbock\inst{\ref{inst1}}
     \and
     O. Krause\inst{\ref{inst1}}
     \and
     H. Dannerbauer\inst{\ref{inst2}}
     \and
     F. Bertoldi\inst{\ref{inst6}}
     \and
     M.-A. Besel\inst{\ref{inst1}}
     \and
     G. de Rosa\inst{\ref{inst1}}
     \and
     X. Fan\inst{\ref{inst5}}
     \and
     M. Haas\inst{\ref{inst4}}
     \and
     D. Hutsemekers\inst{\ref{inst3}}
     \and
     C. Jean\inst{\ref{inst7}}
     \and
     D. Lemke\inst{\ref{inst1}}
     \and
     H.-W. Rix\inst{\ref{inst1}}
     \and
     M. Stickel\inst{\ref{inst1}}}

   \institute{Max-Planck Institut f\"ur Astronomie (MPIA), 
     K\"onigstuhl 17, D-69117 Heidelberg, Germany
     \email{leipski@mpia-hd.mpg.de}\label{inst1}
   \and
   Service d'Astrophysique (SAp)/IRFU/DSM/CEA Saclay - B\^at. 709, 
     91191 Gif-sur-Yvette Cedex, France\label{inst2}
   \and
   Argelander Institut f\"ur Astronomie, Universit\"at Bonn, 
   Auf dem H\"ugel 71, 53121 Bonn, Germany\label{inst6}
   \and
   Steward Observatory, University of Arizona, Tucson, AZ 85721, USA\label{inst5}
   \and
   Astronomisches Institut Ruhr-Universit\"at Bochum, Universit\"atsstra{\ss}e 
     150, 44801 Bochum, Germany\label{inst4}
   \and
   Institut d'Astrophysique et de G\'eophysique, 
     University of Li\`ege, All\'e du 6 Ao\^ut 17, 
     4000 Li\`ege, Belgium\label{inst3}
   \and
   Instituut voor Sterrenkunde, Katholieke Universiteit Leuven, 
   Celestijnenlaan 200B, 3001 Heverlee, Belgium\label{inst7}}
   
\date{}

  \abstract 
{We present {\it Herschel} far-infrared (FIR) observations of two
sub-mm bright quasars at high redshift: SDSS J1148+5251 ($z=6.42$) 
and BR 1202$-$0725 ($z = 4.69$) obtained with the PACS instrument. Both objects are
detected in the PACS photometric bands. The {\it Herschel} measurements
provide additional data points that constrain the FIR spectral energy
distributions (SEDs) of both sources, and they emphasise a broad range
of dust temperatures  in these objects. For
$\lambda_{\rm rest} \lesssim 20$\,$\mu$m, the two SEDs are  very
similar to the average SEDs of quasars at low redshift. In the FIR,
however, both quasars  show excess emission compared to low-$z$
QSO templates, most likely from cold dust powered by vigorous star
formation in the QSO host galaxies. For SDSS  J1148+5251 we detect
another object at 160\,$\mu$m with a distance of
$\sim$\,10\arcsec~from the QSO.  Although no physical connection
between the quasar and this object can be shown with the
available data, it could potentially confuse low-resolution
measurements, thus resulting  in an overestimate of the FIR luminosity
of the $z = 6.42$ quasar. }


   \titlerunning{Far-infrared photometry of two $z>4$ quasars}
   \maketitle

%

\section{Introduction}

The detection of large quantities of cold dust in high redshift
($z>5$) quasars \citep[e.g.][]{ber03,bee06,wan08} implies a substantial
enrichment of the interstellar medium already during the first billion
years after the Big Bang. If the far-infrared (FIR) emission of these
objects is powered by star-formation, then the luminosity and the
temperature of the dust implies star formation rates of up to a few
thousand solar masses per year, possibly indicating the rapid
formation of early galactic bulges.  Recent observations at mid-infrared 
(MIR) wavelengths with {\it Spitzer} demonstrated that the
dust also shows an energetically important hot (T\,$\sim$\,1000\,K)
component in many high-redshift QSOs \citep{hin06,jia06}, with only
few exceptions \citep{jia10}. The similarity of the spectral energy
distributions (SEDs) to those of lower redshift AGN lead to the
conclusion that the general structures characterising local AGN
are already in place at $z\sim6$.  However, for a majority of the sources the (sub)mm
and MIR observations have only measured the tail of the dust emission
spectrum. In order to further
constrain the properties of the dust in these objects, it is
essential to sample the SED as completely as possible. These
measurements can then be used to derive critical parameters such as
the total infrared luminosity and the range of dust temperatures in the 
objects of interest. In addition, the MIR to FIR luminosity ratio may indicate
the relative importance of warm and hot dust (which is predominantly
heated by the AGN) compared to colder dust, which is preferentially
heated by star formation.

The {\it Herschel} key project \citep{pil10} "The Dusty Young
Universe" (PI K.~Meisenheimer) aims to measure the FIR SEDs for all
quasars with $z>5$ that were known at the time of submission of the proposal
(early 2007) with the PACS \citep{pog10} and SPIRE \citep{gri10}
instruments. Together with existing NIR and MIR photometry,
measurements in five {\it Herschel} bands will yield complete infrared
SED coverage of more than two decades in wavelength. Thus, the
properties of the dust and the possible interplay between black-hole
growth and galaxy bulge formation (traced indirectly through the FIR
emission) can be explored early in the evolution of the universe.

\begin{figure*}[t!]
  \centering
  \includegraphics[width=3.5cm]{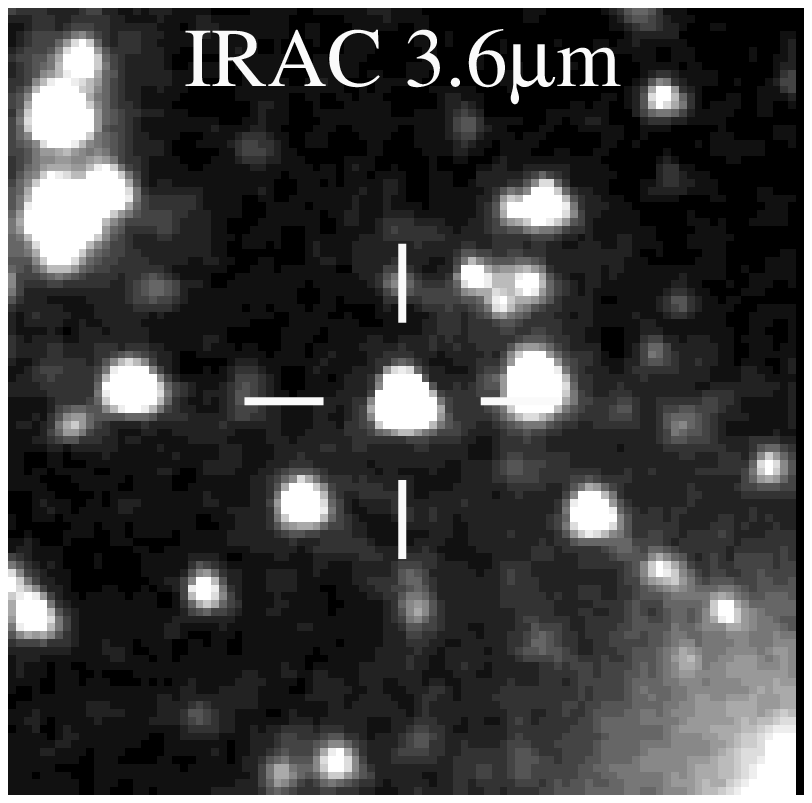}
  \includegraphics[width=3.5cm]{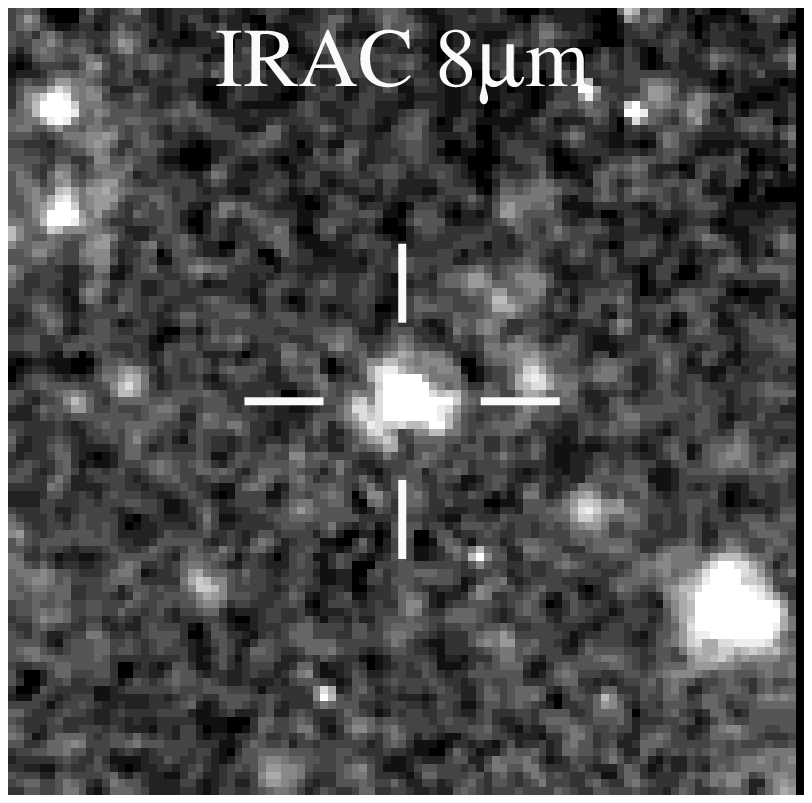}
  \includegraphics[width=3.5cm]{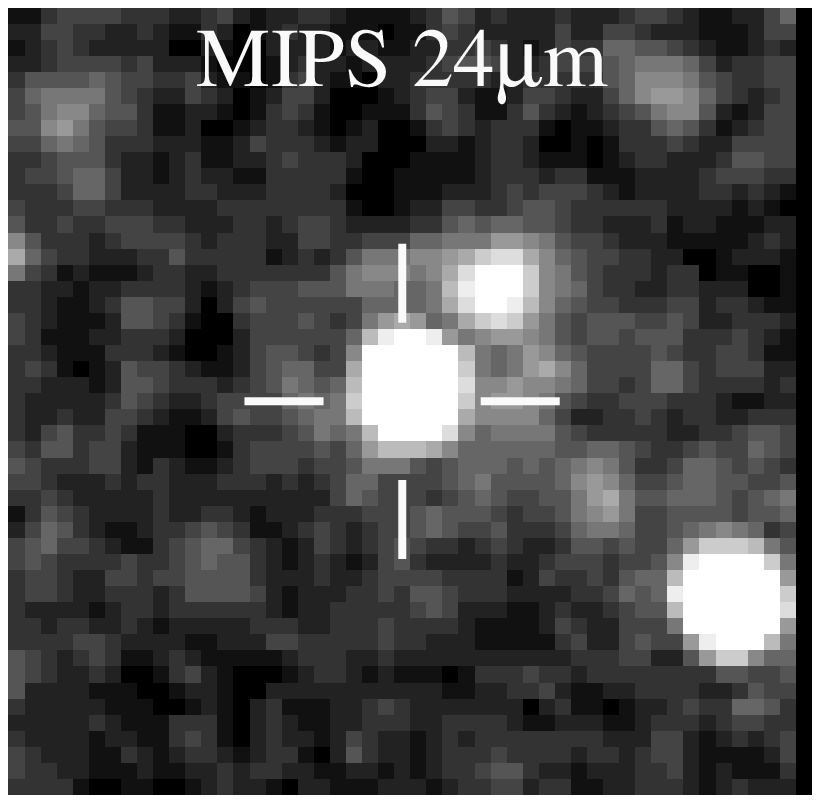}
  \includegraphics[width=3.5cm]{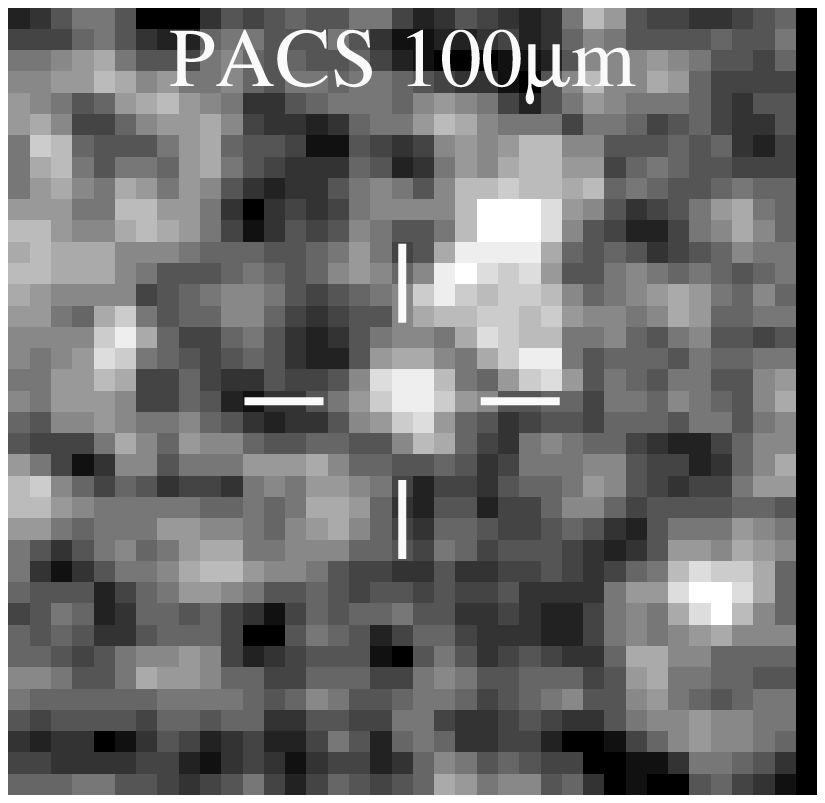}\\
  \includegraphics[width=3.5cm]{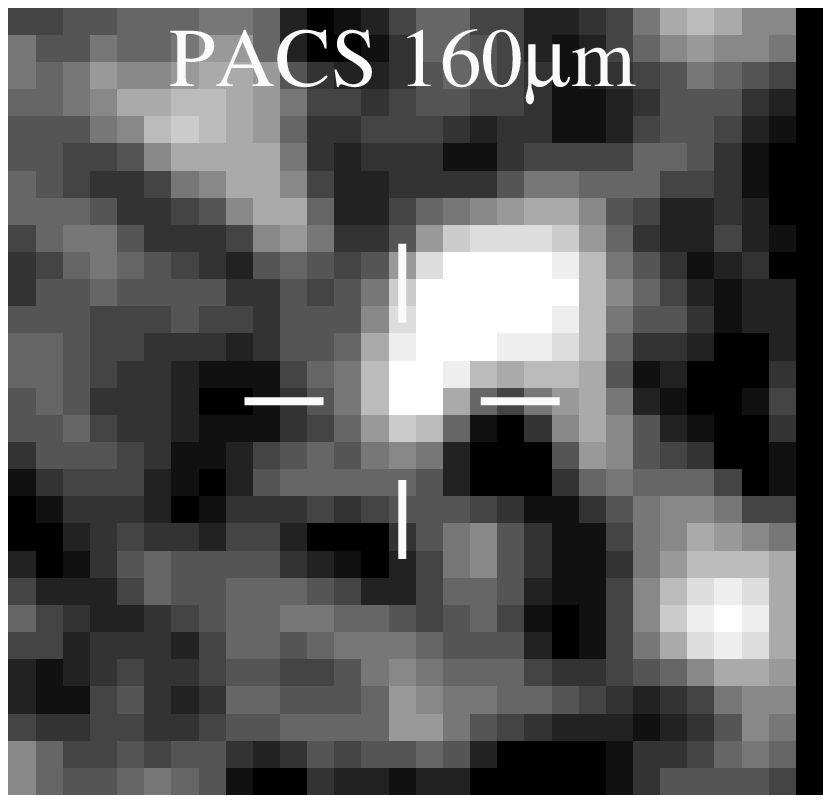}
  \includegraphics[width=3.5cm]{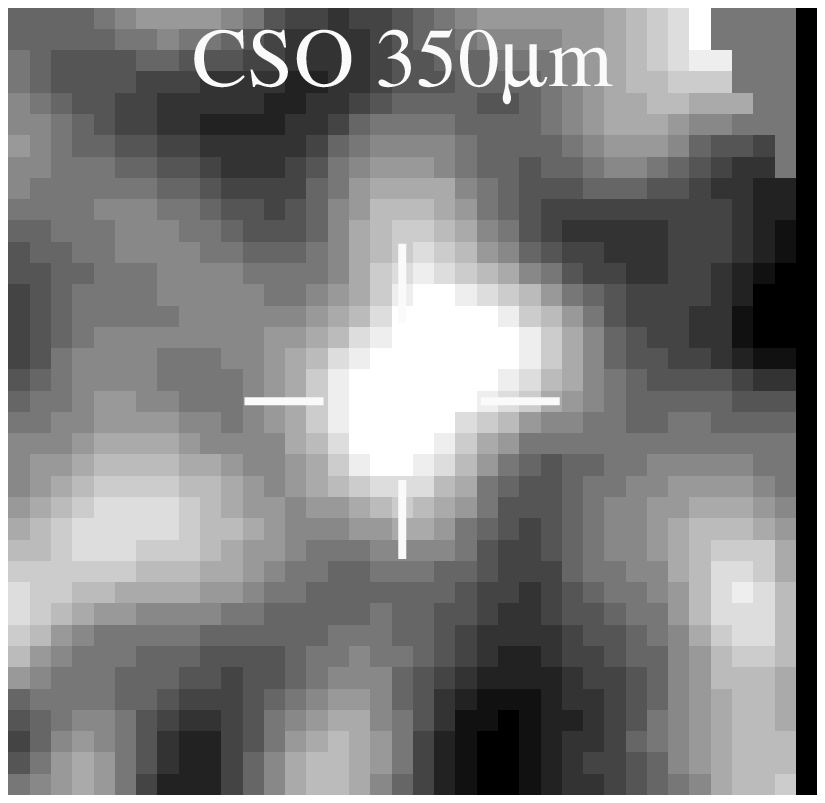}
  \includegraphics[width=3.5cm]{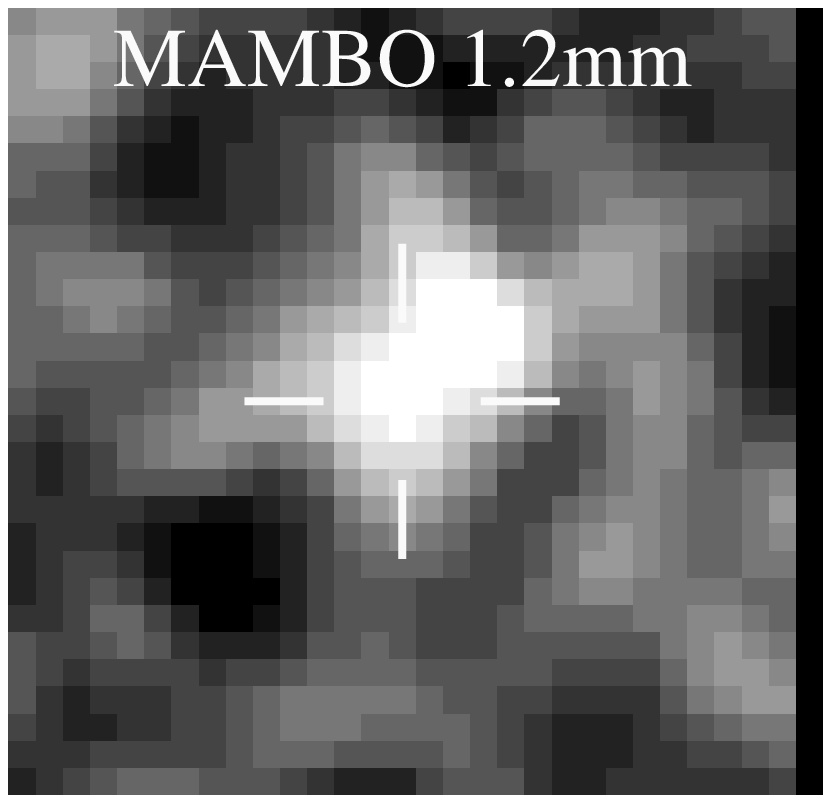}
  \caption{Infrared and (sub)mm images of J1148+5251. All images are 
    60\arcsec~wide and north is up with east to the left.  
    Another nearby source northwest of the QSO (see text) is clearly 
    visible at longer wavelengths.}
 \label{pic_j1148}
\end{figure*}


\section{Observations and data reduction}

During the Science Demonstration Phase (SDP) two of our objects with good ancillary
data were observed with PACS. No SPIRE point-source photometry 
of the sources was obtained during the SDP. 
The objects were selected as SDP targets because they were known to be bright 
at (sub)mm wavelengths. This allows us to study the performance of 
{\it Herschel}/PACS at low flux levels, and the experience
gained here will be beneficial in the further course of the project
when {\it Herschel} will be the only source of FIR data for many object. 
The observations also demonstrate {\it Herschel}'s ability 
to detect dust emission at the highest redshifts.

{\it SDSS J1148+5251.}\quad 
The  quasar SDSS J1148+5251 at $z=6.42$ \citep[][hereafter J1148+5251]{fan03} 
was observed in
PACS scan-map mode using two cross scans in the green (100\,$\mu$m, 
FWHM\,$\sim$\,6\farcs7) 
and in the red (160\,$\mu$m, 
FWHM\,$\sim$\,11\arcsec) band \citep{pog10}, yielding an effective on-source time of
$\sim$1150\,sec in 3810\,sec of total AOR length.  
We employed standard processing
procedures and masked the source location during deglitching and
high-pass filtering. Both scan directions were processed separately
and were then combined into a single map. The fluxes quoted in this paper
were determined using an aperture of 5\farcs0 radius (to avoid
contaminations from neighbouring sources) and the sky was measured
between 30\arcsec~and 35\arcsec. Appropriate aperture corrections were 
included \citep{pog10}\footnote{see also the PACS Photometer release note 
available on the {\it Herschel} Science Centre website {\tt http://herschel.esac.esa.int/}}.  
Due to the faintness of the source
the fluxes measured via aperture photometry in the final maps were
quite sensitive to the parameters chosen during the processing
(i.e. the width of the high-pass filter and the size of the masked
regions). In addition to the scan maps, the source was also observed 
in chop-nod mode ($\sim$2100\,sec on-source time). Despite the
higher noise levels in the chop-nod observations, the source is
securely detected at 100\,$\mu$m (but not at 160\,$\mu$m). 
The scan maps with high-pass filter 
widths of 15 and 20 in the green and red bands, respectively, yield fluxes 
that agree with the chop-nod observations.

{\it BR 1202$-$0725.}\quad 
The $z=4.69$ quasar BR 1202$-$0725 \citep[][hereafter 1202$-$0725]{mcm94} 
was observed in chop-nod mode
in the blue (70\,$\mu$m, 
FWHM\,$\sim$\,5\farcs5) and red band of the PACS instrument for a 
total of 2550\,sec (AOR length) with an on-source time of $\sim$1980\,sec. 
Standard processing steps for chop-nod observations were used in combination
with the latest calibration information. We performed aperture
photometry using apertures with a radius of 7\farcs0 and 10\farcs0 in the
blue and red filters, respectively. The sky was measured at distances between
30\arcsec~and 35\arcsec~away from the source.

\section{Results}

\begin{figure}[b]
  \centering
  \includegraphics[width=3.5cm]{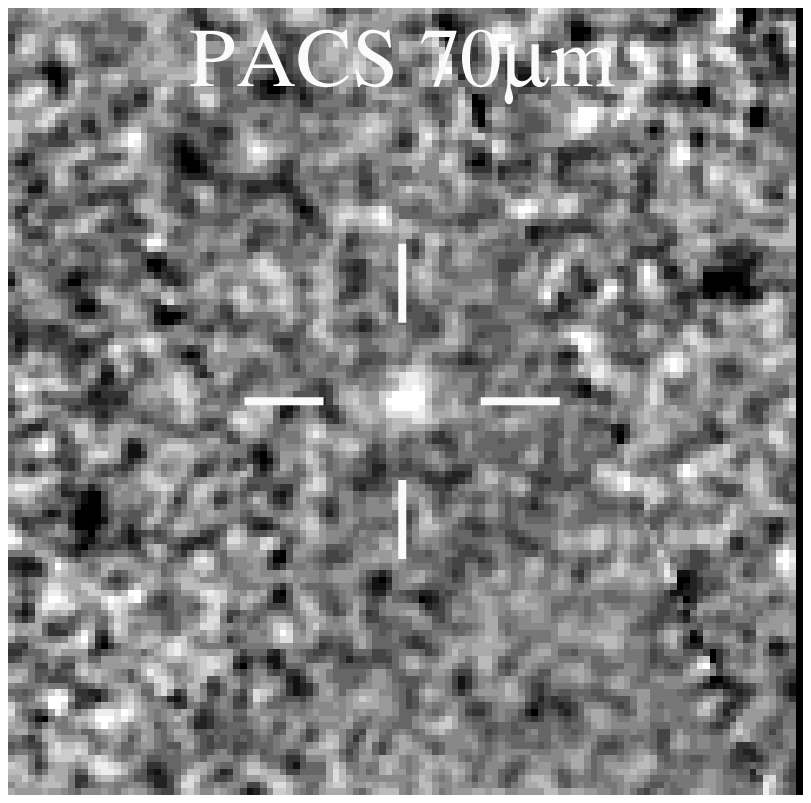}
  \includegraphics[width=3.5cm]{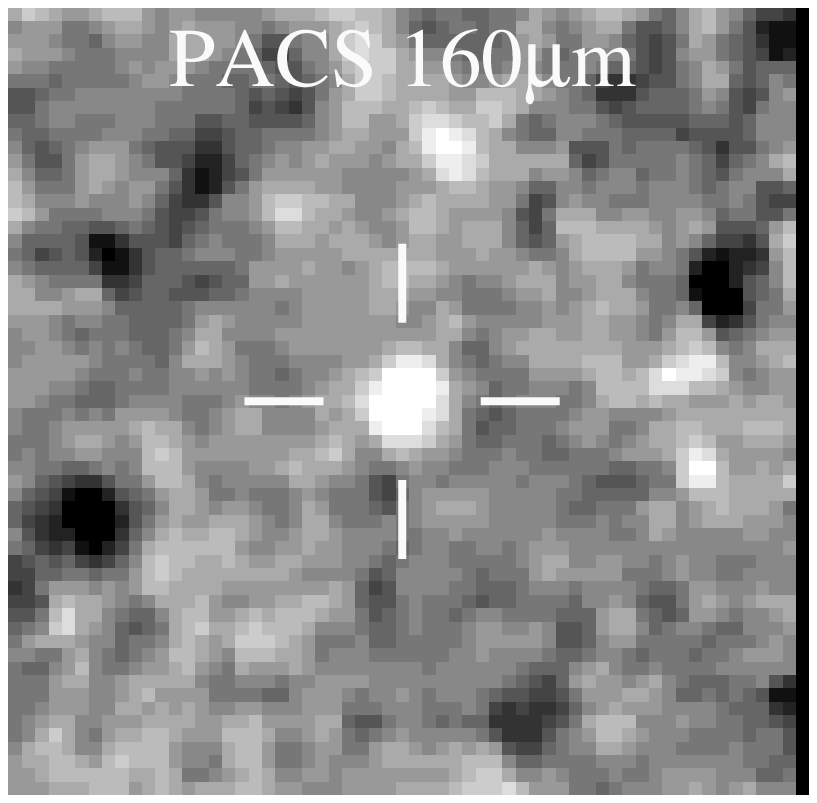}
  \caption{{\it Herschel}/PACS chop-nod images of BR 1202$-$0725. The images are 120\arcsec~wide 
    and north is up with east to the left.} \label{pic_br1202}
\end{figure}

\subsection{J1148+5251}
The QSO J1148+5251 at $z=6.42$ is one of the highest redshift
quasars known to date. It has been detected
previously at several sub-mm \citep{bee06,rob04} and mm
\citep{ber03,rie09} wavelengths.  Under the assumption that the
heating of the cold dust is dominated by young stars, the high FIR
luminosity \citep[$L_{\rm FIR} \sim 2 \times 10^{13}\,L_{\odot}$,][]{bee06} 
translates into a star-formation rate of $\sim
3000\,M_{\odot}\,{\rm yr}^{-1}$.  These high star-formation rates are
corroborated by measurements of the [\ion{C}{II}] emission line at
158\,$\mu$m \citep{mai05,wal09}. Mid-infrared observations
\citep{hin06,jia06} show large amounts of hot dust near 
the $\sim 3 \times 10^{9}\,M_{\odot}$ black hole which accretes close to 
its Eddington limit \citep{wil03,bar03}.

\begin{figure*}[ht!]
  \centering
  \includegraphics[width=6.5cm]{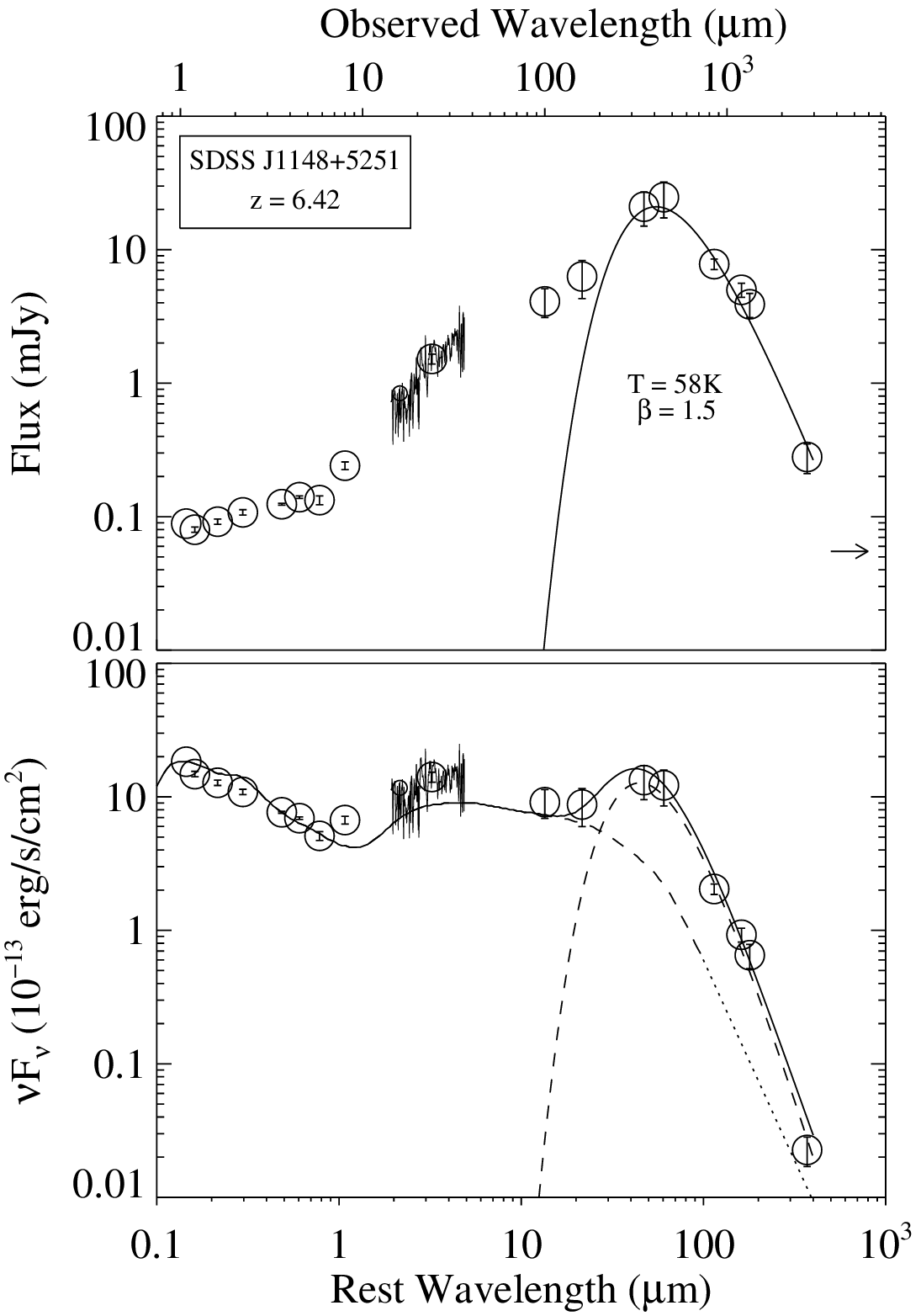}
  \includegraphics[width=6.5cm]{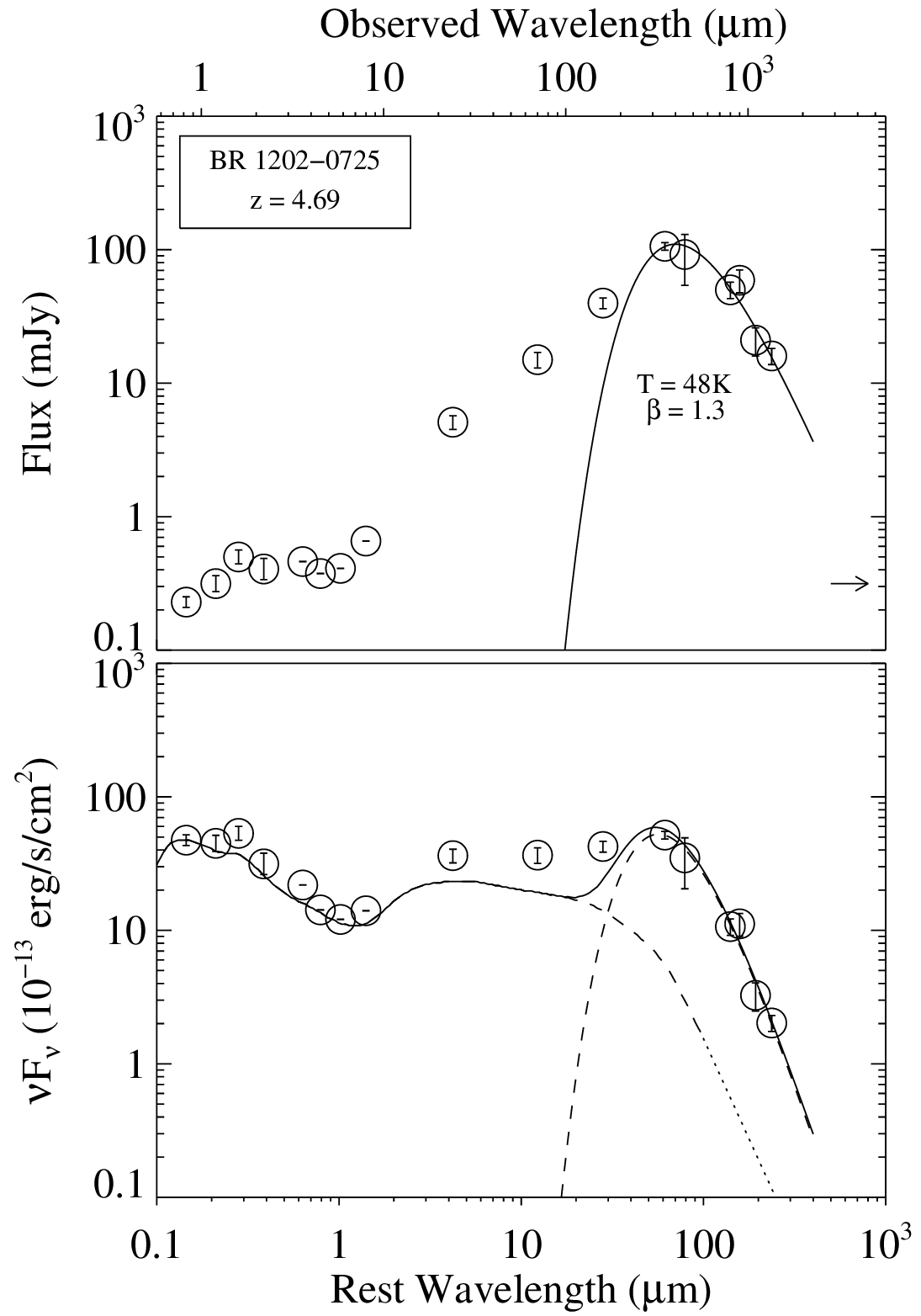}
  \caption{Spectral flux distributions ({\it top}) and spectral energy 
    distributions ({\it bottom}) for J1148+5251 ({\it left}) and for 1202$-$0725 
    ({\it right}). The arrow indicates the level of the 1.4\,GHz
      (observed frame) radio  
    emission (Tab.\,\ref{table1}). In the top plot the solid line shows a single 
    temperature grey body fit to the data at $\lambda_{\rm obs} > 160$\,$\mu$m. In the 
    bottom panels we show as dashed lines the same greybody and also the mean SED for 
    optically-luminous SDSS QSOs \citep{rich06}, scaled to the 
    1450\,\AA~(rest frame) data point. The mean SED was extrapolated for wavelengths 
    longer than 100\,$\mu$m (rest frame) assuming $F_{\nu} \sim {\nu}^2$ (dotted line). 
    The solid line is the sum of the scaled SED and the fitted greybody. Both objects 
    clearly show a large excess of cold dust emission ($\lambda \gtrsim 30$\,$\mu$m, 
    rest frame) compared to the mean QSO template. For J1148+5251 we here also include a 
    {\it Spitzer}/IRS spectrum, which we retrieved from the archive and processed in a 
    standard manner.}
  \label{sed_j1148}
\end{figure*}

Figure\,\ref{pic_j1148} summarises the available multi-wavelength
observations in this field. A careful inspection of the data for
J1148+5251 shows a companion $\sim$\,10\arcsec~to the northwest of
the QSO in the FIR (this is best seen in the 160\,$\mu$m
filter). Indeed, a similar extension is seen in the SHARC\,II
observations \citep{bee06} and in new MAMBO measurements 
\citep[based on][]{ber03} of the source.  
The confusing source is clearly identified in many bands
presented here, and a preliminary analysis of deep HST/ACS imaging of
this field does not reveal an obvious counterpart. Interferometric 
maps at $\sim$1\,mm obtained at IRAM/PdBI \citep{wal09} 
do not show a counterpart for the companion source. However, 
the observations were centred on the QSO, which places the location of 
the companion just outside of the primary beam at that wavelength, thus 
resulting in highly reduced sensitivity. Given the
currently available data on this  source it is clear that further
studies will be needed to reveal its true nature.  We note though
that depending on the shape of its SED this additional source will
affect measurements of J1148+5251 done with large
($\gtrsim$10\arcsec) beams. Aperture photometry  of J1148+5251
yields a flux density at 100 and 160\,$\mu$m of $4.1 \pm 1.0$\,mJy and
$6.3 \pm 2.0$\,mJy, respectively  (see Table\,\ref{table1} for a summary
of all flux measurements, including  literature data). In the red band
we  removed the confusing source via PSF fitting prior to performing
the  photometry.

\subsection{BR 1202$-$0725}
BR 1202$-$0725 is a well studied quasar at $z=4.69$ with strong
detections at sub-mm and mm wavelengths
\citep{isa94,omo96b,ben99,ion06} and in CO line emission
\citep{oht96,omo96a,car02}. At $\sim$4\arcsec~distance to the
northwest of the quasar this source shows a secondary component in the dust
continuum and in CO. A Ly$\alpha$ extension is observed approximately
2\farcs3 northwest of the quasar \citep{pet96,hu96,fon98,ohy04}. We
detected BR 1202$-$0725 at 70\,$\mu$m and 160\,$\mu$m with fluxes of
$15.0\pm2.0$\,mJy and $39.8\pm3.7$\,mJy, respectively
(Fig.~\ref{pic_br1202}), which is consistent with the upper limits
from ISO \citep{lee01}.  In their {\it Spitzer} observations
at 24\,$\mu$m, \citet{hin06} see slightly resolved emission and
perform a two-component PSF fit to isolate the
emission attributable to the QSO. Although our PACS 70\,$\mu$m data
have slightly higher resolution than the MIPS 24\,$\mu$m observations,
we do not detect a secondary component securely, probably due to the 
lower S/N in the 70\,$\mu$m maps compared to the 24\,$\mu$m image. Although 
an apparent elongation may be visible for the red band in Fig.\,\ref{pic_br1202}, a 
two-component fit does not give a significantly better fit to the 
source profile. As we cannot reliably separate the two sources in the PACS 
observations, the quoted flux values refer to the sum of both components.

\section{Discussion and conclusion}

In Fig.~\ref{sed_j1148} we show the spectral flux and energy
distributions of J1148+5251 and 1202$-$0725.  The SEDs of both
high-redshift QSOs appear very similar in shape, and the strong
emission in all infrared  bands shows dust at a wide range of
temperatures. This is particularly  emphasised by the {\it Herschel}
photometry, which fills the gap between the previously available MIR
and sub-mm photometry.  The shape of the SED also implies that
large amounts of dust are present in the host galaxies already at high
redshift and that the dust may be distributed on a wide range of
scales: NIR emission from hot dust close to the nucleus
\citep[e.g.][]{hin06,jia06,jia10}, warm MIR dust (T\,$\sim$\,few
hundred K) on intermediate scales (or partly shielded), and colder
dust in the FIR possibly distributed throughout the host galaxy.  
Using the new {\it Herschel} photometry to determine
total infrared luminosities for these two objects (integrating the
SEDs between 1 and 200\,$\mu$m, rest frame) yields  $L_{\rm FIR} = 8.1 \times
10^{13}\,L_{\odot}$ for J1148+5251 and a total $L_{\rm FIR} = 3.7 \times
10^{14}\,L_{\odot}$ for 1202$-$0725.

The inspection of the spectral energy distributions  in
Fig.\,~\ref{sed_j1148} shows that in the UV/optical and  NIR/MIR (see
below), where the emission is dominated by the  active nucleus,
the SEDs of the high-$z$ sources match the template SEDs  from lower
redshift AGN \citep{rich06}  reasonably well. However, while the
low-$z$ infrared SEDs show only one broad peak at short MIR
wavelengths, we observe a second peak with considerable excess
emission at FIR wavelengths in our targets.  Greybody fits to the
data at $\lambda > 160$\,$\mu$m yield a  temperature of
$\sim$\,$50-60$\,K for the FIR dust emission, which is likely powered
by the vigorous star formation \citep[see
also][]{ber03,bee06,wan08}. These large contributions to the FIR
emission,  which can presumably  be attributed to ongoing star
formation suggesting rapid bulge build-up in the  host galaxies at
high redshift, are missing in most lower redshift AGN. A  combination
of the mean QSO SED and the single temperature greybody representing
the contributions from star formation is able to match the observed
photometry.  We note however that both objects presented here
were selected as  SDP targets because of their known strong
FIR/sub(mm) emission. In  principle, they could represent a small
fraction of strongly star-forming objects,  while the majority of the
high-$z$ QSOs may lack such a powerful FIR component.  An analogue for
this situation may be found in the (mostly local) PG quasars
\citep{haas03}: when compared to the high-$z$ objects, many of the PG
sources  show a similar mismatch in the FIR as seen for the SDSS QSO
template, but a small  number of PG quasars reveal a comparable FIR
excess as seen for the high-redshift QSOs.  Alternatively, stronger
FIR emission from bulge build-up via star formation may be  more
common at high $z$ than at low $z$, and we plan to explore these
questions once data  for a greater number of high-redshift objects 
become available.

Both QSOs show more flux at NIR and MIR wavelengths than we would
expect from the local templates. For 1202$-$0725 this may be
understood to be due to the inclusion of the companion (which is at
the same redshift as the quasar;  \citealt{omo96a}, \citealt{car02})
and which contributes roughly half of  the flux in many infrared and
(sub)mm bands  \citep[e.g.][]{omo96a,hin06,ion06}. For 
J1148+5251 the situation is somewhat different, because the QSO is
clearly the dominating  source in flux at e.g. 8 and 24\,$\mu$m
(observed, Fig.~\ref{pic_j1148}). However, the exceptionally
luminous  black hole in this QSO accretes close to its Eddington limit,
which could possibly  lead to a larger fraction of dust being heated
to high temperatures, resulting  in increased NIR emission.

For both objects a secondary component may contribute to the measured
flux densities at FIR wavelengths, potentially leading to  an
overestimate of the FIR luminosity of the QSO itself.

\begin{table}[h]
\caption{Multi-wavelength data. All fluxes are given in mJy.}      
\label{table1}     
\begin{center}       
\begin{tabular}{ccccc} 
\hline\hline             
$\lambda_{\rm obs}$ & SDSS J1148+5251        & refs & BR 1202$-$0725 & refs \\ 
in $\mu$m          & $z=6.42$               &            & $z = 4.69$     & \\
\hline                       
  0.1450\tablefootmark{a} & 0.0887          &  1  & 0.23$\pm$0.02   & 11 \\  
  1.2                     & 0.080$\pm$0.004\tablefootmark{b} &  1  & 0.32$\pm$0.04\tablefootmark{b}   & 12   \\
  1.6                     & 0.092$\pm$0.004\tablefootmark{b} &  2  & 0.50$\pm$0.06\tablefootmark{b}   & 12    \\
  2.2                     &  0.11$\pm$0.005\tablefootmark{b} &  2  & 0.41$\pm$0.08\tablefootmark{b}   & 12   \\
  3.6                     & 0.124$\pm$0.002 &  3  & 0.461$\pm$0.001 &  4    \\ 
  4.5                     & 0.140$\pm$0.003 &  3  & 0.375$\pm$0.001 &  4    \\ 
  5.8                     & 0.133$\pm$0.010 &  3  & 0.410$\pm$0.001 &  4    \\ 
  8.0                     & 0.241$\pm$0.016 &  3  & 0.657$\pm$0.002 &  4    \\ 
 16.0                     & 0.84           &  3  & --             &--      \\ 
 24.0                     & 1.520$\pm$0.130 &  4  & 5.1$\pm$0.6\tablefootmark{c}     &  5   \\ 
 70.0                     & --             & -- & 15.0$\pm$2.0\tablefootmark{c}    &  5   \\ 
100.0                     & 4.1$\pm$1.0     &  5  & --             &--     \\ 
160.0                     & 6.3$\pm$2.0     &  5  & 39.8$\pm$3.7\tablefootmark{c} &  5   \\ 
350.0                     &  21$\pm$6       &  6  & 106$\pm$7\tablefootmark{c}       & 13     \\ 
450.0                     & 24.7$\pm$7.4    &  7  & 92$\pm$38\tablefootmark{c}       & 14     \\ 
800.0                     & --             & -- & 50$\pm$7\tablefootmark{c}        & 14    \\ 
850.0                     & 7.9$\pm$0.7     &  7  & --            & --     \\ 
900.0                     & --             & -- & 59$\pm$11\tablefootmark{c}       & 15     \\ 
1100.0                    & --             & -- & 21$\pm$5\tablefootmark{c}        &14 \\
1200.0                    & 5.0$\pm$0.6     &  8  & --             &-- \\
1330.0                    & 3.9$\pm$0.8     &  9  & --             &-- \\
1350.0                    & --             & -- & 16.0$\pm$2.2\tablefootmark{c}    & 16 \\
2750.0                    & 0.28$\pm$0.07   &  9  & --             &-- \\
1.4\,GHz                  & 0.055$\pm$0.8   &  10 & 0.315$\pm$0.080\tablefootmark{c} & 17 \\
\hline                       
\end{tabular}
\end{center}       
\tablefoottext{a}{Refers to the 1450\,\AA~flux in the rest frame of the source.}\\
\tablefoottext{b}{Calculated using the zero points presented in \citet{coh03}.}\\
\tablefoottext{c}{Total flux including the companion.}
\tablebib{(1)~\citealt{fan03}; (2) \citealt{iwa04}; (3) \citealt{jia06}; 
      (4) \citealt{hin06}; (5) this work; (6) \citealt{bee06}; (7) \citealt{rob04}; 
      (8) \citealt{ber03}; (9) \citealt{rie09}; (10) \citealt{car04}; 
      (11) \citealt{sto96}; (12) \citealt{skr06}; (13) \citealt{ben99}; (14) \citealt{isa94};
      (15) \citealt{ion06}; (16) \citealt{omo96a}; (17) \citealt{yun00}.}
\end{table}

\begin{acknowledgements}

PACS has been developed by a consortium of institutes led by MPE
(Germany) and including UVIE (Austria); KU Leuven, CSL, IMEC
(Belgium); CEA, LAM (France); MPIA (Germany); INAFIFSI/ OAA/OAP/OAT,
LENS, SISSA (Italy); IAC (Spain). This development has been supported
by the funding agencies BMVIT (Austria), ESA-PRODEX (Belgium),
CEA/CNES (France), DLR (Germany), ASI/INAF (Italy), and CICYT/MCYT
(Spain). M.H. is supported by the Nordrhein-Westf\"alische Akademie der 
  Wissenschaften und der K\"unste. D.H. is supported by F.R.S.-FNRS
(Belgium). We thank the referee for useful comments that helped to improve 
the paper.

\end{acknowledgements}

\end{document}